\documentclass[11pt,preprint]{aastex}
\usepackage{longtable}

\newcommand{\etal}{{\it et~al.}}

\begin{document}

\title{A family-based method of quantifying NEOWISE diameter errors}

\author{Joseph R. Masiero\altaffilmark{1}, A.K. Mainzer\altaffilmark{1}, E.L. Wright\altaffilmark{2}}

\altaffiltext{1}{Jet Propulsion Laboratory/California Institute of Technology, 4800 Oak Grove Dr., MS 183-301, Pasadena, CA 91109, USA, {\it Joseph.Masiero@jpl.nasa.gov}}
\altaffiltext{2}{University of California, Los Angeles, CA, 90095}

\begin{abstract}
Quantifying the accuracy with which physical properties of asteroids
can be determined from thermal modeling is critical to measuring the
impact of infrared data on our understanding of asteroids.  Previous
work \citep{mainzer11cal} has used independently-derived diameters
(from asteroid radar, occultations, and spacecraft visits) to test the
accuracy of the NEOWISE diameter determinations.  Here, we present a
new and different method for bounding the actual NEOWISE diameter
errors in the Main Belt based on our knowledge of the albedos of
asteroid families.  We show the $1\sigma$ relative diameter error for
the Main Belt population must be less than $17.5\%$ for the vast
majority of objects. For a typical uncertainty on $H$ magnitude of
$0.2~$mag, the relative error on diameter for the population would be
$\sim10\%$.

\end{abstract}

\section{Introduction}

Asteroid families are composed of objects formed from the catastrophic
breakup of a single, larger parent body
\citep{masieroAIV,michelAIV,nesvornyAIV}.  For objects that have not
undergone partial- or total-differentiation, it is reasonable to
assume that the original parent had a macroscopically homogeneous
composition, and that each member of the family represents a single,
independent sample of this composition \citep{michelAIV}.  Similarly,
as asteroid family members are all the same age (with the caveat that
a small fraction of members may have undergone later impact events)
and reside in nearly the same location in the Solar system, we can
assume that any evolution of the surface (e.g. due to space
weathering) affects all members in the same fashion
\citep{nesvornyAIV}.  Thus, by measuring a particular property for
many family members we can better quantify the true value of the
original parameter that is being sampled.  In this work, we use the
average and variance of the albedos measured for all members of an
asteroid family ($\widehat{p_V}$, $\sigma_{\widehat{p_V}}^2$) as an
estimator for the true albedo ($p_V$).

\section{Relationship between errors on diameter, absolute magnitude, and albedo}

Asteroid albedo ($p_V$) and diameter ($D$) are related through the
absolute magnitude ($H$) by the equation \citep[][]{harris02}:

\begin{equation}
D_{km} = 1329~\frac{10^{-H/5}}{\sqrt{p_V}}
\label{eq.orig}
\end{equation}

The $H$ magnitude is derived from a fit to the observed visible light
apparent magnitudes, often using an assumed phase curve slope
parameter $G$ \citep{bowell89}.  Diameter is determined through a
thermal model fit to measured infrared emitted fluxes.  Albedo is then
inferred from these two parameters. The uncertainty on an albedo
measurement will therefore derive from the uncertainties on the
diameter as measured by e.g. NEOWISE, and the uncertainty on the $H$
magnitude.  The $H$ magnitude uncertainty is driven by the uncertainty
in the apparent magnitude measurements from ground-based telescopes as
well as by the assumed phase curve slope parameter $G$.

By converting Equation~\ref{eq.orig} from absolute magnitude to an effective flux ($F_H$) and
propagating the uncertainty ($\sigma$) on each of these parameters, we
find that:

\begin{equation}
\sqrt{p_V} = C_1 \frac{\sqrt{2.511^{-H}}}{D}
\end{equation}

\begin{equation}
p_V = C_2 \frac{F_H}{D^2}
\end{equation}

\begin{equation}
\left(\frac{\sigma_{p_V}}{p_V}\right)^2 = \left(\frac{\sigma_{F_H}}{F_H}\right)^2 + \left(\frac{2\sigma_D}{D}\right)^2 - 2 \frac{\sigma_{D F_H}}{D F_H}  
\label{eq.fullerror}
\end{equation}

where where $C_1$ and $C_2$ are constants and $\sigma_{D F_H}$ is the
covariance between diameter and the $H$ magnitude. As $H$ is based on
reflected visible light measurements and $D$ is based on emitted
thermal infrared fluxes, the measurements are independent and
$\sigma_{D F_H}=0$.

In the case where we have an independent measurement of the
  fractional error on albedo, we can use the relationship in
  Equation~\ref{eq.fullerror} to determine the maximum possible error
  on either diameter or $H$ magnitude by assuming the error on the
  other is zero.  This is of course unphysical, but allows us to
  determine the worst-case scenario for the errors on one parameter.
  Thus, if the $H$ magnitude is perfectly known ($\sigma_{F_H}=0$),
the error on diameter and albedo will be directly correlated. For
non-zero values of $\sigma_{F_H}$, the albedo error is a combination
of the $H$ and $D$ errors.  This therefore sets an upper limit on the
diameter error from a measurement of the albedo error of:

\begin{equation}
\frac{\sigma_D}{D} < 0.5 \frac{\sigma_{p_V}}{p_V}.
\end{equation}

Therefore, a quantification of the fractional uncertainty on an albedo
determination for a population ($\sigma_{\widehat{p_V}}$) sets an
upper limit on the fractional uncertainty for the diameter
determination of that population.

We note that propagating the error using Equation~\ref{eq.orig} as written
will result in a term that includes the covariance of albedo and
absolute magnitude. As all of the albedos used for this study were
determined through thermal modeling and visible light $H$ magnitude
fitting, albedo and $H$ are not independent, and the covariance
between these two parameters is large and significant.

\section{Asteroid families as tests of albedo and diameter errors}

With asteroid families, we effectively have a series of independent
measurements of the same surface.  As objects were not known {\it a
  priori} to be a family member when they were observed by NEOWISE
\citep{mainzer11} or when physical properties were determined
\citep{masiero11}, and because the objects were detected over a wide
range of observing geometries and sky positions, there are no
correlated systematic errors that can affect one family without
affecting the entire catalog.  Thus, families act as a method of
independent verification of the quality of the thermal model fits.
The albedos of objects with common spectral taxonomy could also be
used in a similar fashion \citep[e.g.][]{mainzer11tax}.  However, it
is conceivable that a selection effect mechanism could induce a
separate systematic error, as albedo and taxonomy both are driven by
optical reflectance properties.  Family membership, conversely, is
established by orbital dynamics \citep{nesvornyAIV}, in particular the
time-integrated proper orbital elements.  For this work, we use the
most recent asteroid family list from the Planetary Data System
\citep[PDS,][]{nesvornyPDS}.

It is important to note that family membership lists are now often
filtered by albedo following the release of the NEOWISE physical property
measurements to remove interlopers \citep[e.g.][]{masiero13,milani14}.
This is because the Main Belt has a bimodal albedo distribution
\citep{masiero11} and interlopers from the other compositional complex
can be easily identified.  Although this could potentially result in a
decrease in the scatter in the albedo measurements, in practice it is
not applied with this level of precision and instead acts only as a
coarse cut to remove objects with grossly inconsistent compositions.
The principal outlier in this process is the Nysa-Polana family
complex, which is known to consist of at least two overlapping
families of different compositions
\citep{cellino02,masiero11,masieroAIV}, but is still listed as a
single entry in PDS.  For this reason, we have removed the Nysa-Polana
complex from consideration in this work.

The results of Anderson-Darling tests of normality for the albedo
distributions of each family show that the null hypothesis of
normality cannot be rejected at the $p=0.05$ probability level for
one-third of the families. We note that the Anderson-Darling
  test is particularly sensitive to differences in the tails of the
  distribution from a normal distribution. For the families where the
  null hypothesis could be rejected, comparison of the cumulative
distribution functions of the albedos and the best-fit Gaussian show
the albedo distributions typically are narrower than the low wing but
broader than the high wing of the Gaussian.  This is likely due to the
fact that some of the uncertainties on the parameters that contribute
to albedo do not follow a normal distribution.  For example, the
uncertainty on apparent magnitude due to unknown rotational phase will
follow a sinusoidal distribution, and light curve maxima will be
better measured than minima in a magnitude-limited survey.  This would
result in an overestimate of the brightness, and thus a skew to higher
albedos, as is observed.  Thus, while the albedos of some families do
not follow a precise normal distribution, the width of the best-fit
Gaussian provides a useful metric for characterizing the systematics
encountered in asteroid thermal modeling.

\section{Results}

\begin{figure}[ht]
\begin{center}
  \includegraphics[scale=0.6]{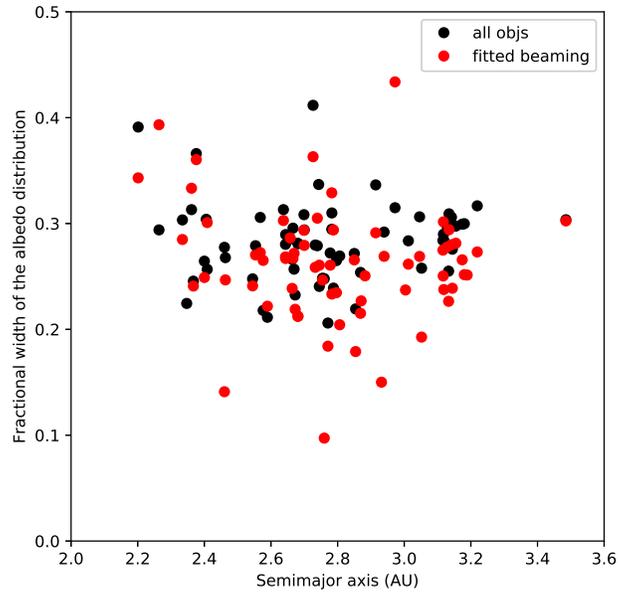}
  \protect\caption{Fractional width of the albedo distribution
    ($w=\frac{\sigma_{\widehat{p_V}}}{\widehat{p_V}}$) for the
    65 families with more than 50 measured albedos (black), as a
    function of distance to the Sun.  Red points show the fractional
    width of only objects with fitted beaming parameters, for 68
    families with more than 20 measured albedos.}
\label{fig.siga}
\end{center}
\end{figure}

\begin{figure}[ht]
\begin{center}
\includegraphics[scale=0.6]{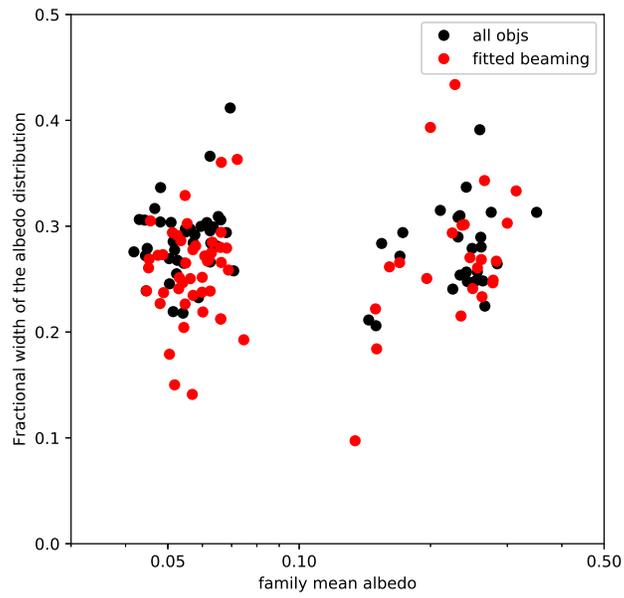} \protect\caption{The same
  as Fig~\ref{fig.siga} but compared to the mean albedo of the family
  ($\widehat{p_V}$).}
\label{fig.sigalb}
\end{center}
\end{figure}

We show in Figures~\ref{fig.siga} and \ref{fig.sigalb} the fractional
width of the albedo distributions for asteroid families as a function
of distance from the sun, and mean family albedo.  This fractional
width is the $1\sigma$ spread of the best-fit Gaussian to the albedo
distribution divided by the mean albedo of the family
($w=\frac{\sigma_{\widehat{p_V}}}{\widehat{p_V}}$).  We plot
each family that has more than 50 albedo measurements in the NEOWISE
data set \citep{mainzer16}.  In total we use $35,468$ asteroids for
this analysis.  

We separately analyze only those objects that had a fitted beaming
parameter in the NEOWISE data set.  The beaming parameter of the
NEATM thermal model was allowed to vary when an observation included two
thermally dominated channels, and accounts for uncertainties in the
model and physical properties of the object \citep{mainzer11cal}.  Of
the $137,014$ unique MBAs with published NEOWISE-derived parameters,
$65,323$ have fitted beaming values, while the others were assumed to
be a fixed value during fitting \citep{masiero11,masiero14}.  As they
are a smaller population, we calculate the fractional width for all
families with more than 20 albedo measurements from fitted-beaming
diameter determinations.  In total $20,150$ asteroids with fitted
beaming were used here.

We note that this is a parallel analysis to what is published in
\citet{masieroAIV}.  There, the authors quote the standard deviation
of the mean of the albedo distribution, which represents the accuracy
with which the mean albedo is known, and decreases linearly with the
number of objects measured.  The spread of the distribution, being the
width of the Gaussian fit, should remain constant for any peak value
as long as the sample size is sufficient.

We list our measured $w$ values for all family members used
  and only those family members with fitted beaming values in
  Table~\ref{tab.w}, along with the numbers of objects used for each
  family.  Our measurements show that the typical spread for family
albedos is $20\% - 35\%$, with a characteristic value of $27\%$.  All
but two families have a fractional width of $w<40\%$.  This means that
the maximum possible uncertainty on NEOWISE diameter measurements for
family members is $20\%$, with the vast majority below $\sim 17.5\%$.
The true diameter uncertainty is significantly lower as this assumes
that the $H$ magnitudes are perfectly known, whereas in reality $H$
magnitudes are known to have errors of order $0.2~$mag, or $\sim 20\%$
in flux \citep{pravec12,veres15}.  If we consider only the objects
with fitted beaming values, we find that the $w$ measurements decrease
slightly, as does the characteristic spread for all families.  This
also confirms that the assumed beaming parameters used for fits with
only a single thermal band are appropriate for the population.

Previous works \citep[e.g.][]{masiero11} have used the log of
  the albedo as a primary diagnostic quantity.  As a test, we have
  performed a parallel analysis of the distribution of $\log p_V$ for
  each family.  We find a comparable rate of success and failure of
  the Anderson-Darling normality test for the log-distributions, and
  the measured widths of the best-fit gaussians match those in
  Table~\ref{tab.w} when translated to linear-space. As the use of
  logarithms adds another layer of complexity to the
  error-propagation, we prefer to use the linear-albedo distributions
  here, but both produce comparable results.

Errors on diameter determination come from two primary
  sources: the measurement uncertainties on the fluxes that are used
  for thermal modeling, and difference between the simplified model of
  the asteroid's thermal behavior and the actual thermal behavior of
  the surface. A single measurement at SNR$=5$ will have a $20\%$
  uncertainty on infrared flux which translates to a $20\%$
  uncertainty on surface area, and thus $\sim10\%$ uncertainty
  diameter. As the number of detection and SNR of detections
  increases, this component of the error diminishes.  The NEOWISE
  object detection pipeline required five observations with SNR$>4.5$,
  for a total SNR$\sim10$.  Therefore, the majority of the uncertainty
  on diameter derives from the differences between the thermal model
  used and the reality of the asteroid surface, and that is the
  uncertainty we are constraining with the test described here.

\clearpage

\section{Conclusions}

The diameter errors for NEOWISE as told by asteroid families are
bounded to be less than $17.5\%$ for the vast majority of objects
assuming perfect knowledge of the $H$ magnitude, and
certainly lower when the uncertainty on the $H$ magnitude is taken
into account.  If the assumption that the composition of family
members is homogeneous is not correct, then natural variations will
inflate the width of the albedo distribution, and this will further
lower the upper limit on diameter accuracy.  A $1\sigma$ uncertainty
of $10\%$ in diameter and $0.2~$mags in $H$ magnitude will result in a
$28\%$ uncertainty on the albedo determination, which corresponds to
the average dispersion we measure for all families.  An
over-estimation of diameter uncertainty from a theoretical case can
occur when the assumptions about the input physical parameters are
allowed to vary beyond values supported by reality.  Mathematical
solutions to many equations are possible that are contradicted by data
from the physical world.  We therefore use data-based interpretations
when available to quantify the real-world NEOWISE modeling
uncertainties.

\textbf{Acknowledgments:} We thank both referees for comments that led
to clarifications, refinements, and improvements to this manuscript,
as well as the editor Melissa McGrath for helping shepherd this
through the referee process.  This research was carried out at the Jet
Propulsion Laboratory, California Institute of Technology, under a
contract with the National Aeronautics and Space Administration.  This
publication makes use of data products from the Wide-field Infrared
Survey Explorer, which is a joint project of the University of
California, Los Angeles, and the Jet Propulsion Laboratory/California
Institute of Technology, funded by the National Aeronautics and Space
Administration.  This publication also makes use of data products from
NEOWISE, which is a project of the Jet Propulsion
Laboratory/California Institute of Technology, funded by the Planetary
Science Division of the National Aeronautics and Space Administration.

\clearpage

\scriptsize{
\begin{longtable}{clcrcr}
  \caption{Fractional width of the albedo distribution ($w$) measured
  for families from all members with albedo measurements, and only
  those with fitted beaming values ($w_{beam}$). The number of family
  members used is listed ($N$) as well as the family name and ID
  number from the PDS table \citep{nesvornyPDS}. Families with
  insufficient numbers of objects ($N<50$ for all members or $N<20$
  fitted beaming values) are listed as `n/a'.}\\
Family ID&          Name &  $w$ &  $N$ & $w_{beam}$ & $N_{beam}$\\
\endhead
401   &         Vesta & 0.31 & 1853 & 0.33 &   542\\
402   &         Flora & 0.39 & 3008 & 0.34 &  1204\\
403   &    Baptistina & 0.29 &  646 & 0.39 &   233\\
404   &      Massalia & 0.26 &  246 & 0.30 &    33\\
406   &       Erigone & 0.25 &  806 & 0.24 &   404\\
407   &      Clarissa & 0.30 &   67 &  n/a &   n/a\\
408   &     Sulamitis & 0.27 &  170 & 0.25 &    97\\
410   &       Euterpe & 0.22 &   57 &  n/a &   n/a\\
413   &          Klio & 0.30 &  320 & 0.28 &   197\\
414   &      Chimaera & 0.28 &   72 & 0.14 &    49\\
415   &      Chaldaea & 0.37 &  126 & 0.36 &    86\\
501   &          Juno & 0.26 &  218 & 0.27 &    76\\
502   &       Eunomia & 0.28 & 2069 & 0.27 &  1166\\
504   &       Nemesis & 0.29 &  396 & 0.28 &   165\\
505   &        Adeona & 0.23 & 1335 & 0.22 &   896\\
506   &         Maria & 0.28 &  954 & 0.27 &   425\\
507   &         Padua & 0.28 &  582 & 0.26 &   317\\
509   &       Chloris & 0.41 &  133 & 0.36 &    83\\
510   &          Misa & 0.29 &  310 & 0.29 &   149\\
512   &          Dora & 0.26 &  803 & 0.23 &   543\\
513   &        Merxia & 0.24 &   99 & 0.26 &    27\\
514   &         Agnia & 0.31 &  114 & 0.23 &    44\\
515   &        Astrid & 0.24 &  220 & 0.29 &    68\\
516   &        Gefion & 0.25 &  640 & 0.25 &   283\\
517   &         Konig & 0.31 &  235 & 0.27 &   114\\
518   &        Rafita & 0.25 &  242 & 0.24 &    81\\
519   &   Hoffmeister & 0.27 &  967 & 0.26 &   440\\
522   &           Ino & 0.34 &   81 &  n/a &   n/a\\
528   &      Leonidas & 0.28 &   68 & 0.21 &    50\\
529   &       Vibilia & 0.28 &   68 & 0.21 &    50\\
530   &         Phaeo & 0.29 &  136 & 0.33 &    93\\
531   &      Mitidika & 0.27 &  635 & 0.24 &   446\\
532   &         Henan & 0.31 &  144 & 0.29 &    49\\
533   &         Hanna & 0.27 &  116 & 0.20 &    54\\
534   &         Karma & 0.22 &   77 & 0.27 &    45\\
535   &          Witt & 0.25 &   75 &  n/a &   n/a\\
537   &      Watsonia &  n/a &  n/a & 0.10 &    32\\
541   &      Postrema & 0.28 &   65 & 0.31 &    43\\
601   &        Hygiea & 0.31 & 2067 & 0.28 &  1396\\
602   &        Themis & 0.31 & 2186 & 0.29 &  1624\\
603   &        Sylvia & 0.30 &  151 & 0.30 &    99\\
604   &      Meliboea & 0.28 &  270 & 0.24 &   205\\
605   &       Koronis & 0.25 & 1053 & 0.22 &   587\\
606   &           Eos & 0.28 & 3726 & 0.26 &  2197\\
607   &          Emma & 0.31 &  310 & 0.27 &   220\\
608   &      Brasilia & 0.27 &  115 & 0.27 &    41\\
609   &       Veritas & 0.30 &  678 & 0.27 &   410\\
611   &         Naema & 0.29 &  231 & 0.27 &   156\\
612   &        Tirela & 0.29 &  282 & 0.30 &   116\\
613   &     Lixiaohua & 0.28 &  510 & 0.24 &   367\\
614   &     Telramund & 0.31 &   83 & 0.43 &    22\\
617   &     Theobalda & 0.30 &  172 & 0.25 &   122\\
618   &      Terentia &  n/a &  n/a & 0.15 &    23\\
622   &   Terpsichore & 0.22 &   81 & 0.18 &    54\\
623   &     Fringilla & 0.34 &  102 & 0.29 &    81\\
625   &      Yakovlev &  n/a &  n/a & 0.23 &    29\\
626   &   Sanmarcello &  n/a &  n/a & 0.25 &    23\\
630   &         Aegle & 0.26 &   71 & 0.19 &    58\\
631   &        Ursula & 0.30 &  891 & 0.28 &   720\\
632   &      Elfriede &  n/a &  n/a & 0.25 &    29\\
638   &       Croatia & 0.26 &  104 & 0.23 &    68\\
641   &       Juliana &  n/a &  n/a & 0.24 &    24\\
701   &       Phocaea & 0.26 &  740 & 0.25 &   344\\
801   &        Pallas & 0.21 &   64 & 0.18 &    47\\
803   &         Hansa & 0.29 &  354 & 0.27 &   123\\
804   &      Gersuind & 0.21 &  138 & 0.22 &    51\\
805   &     Barcelona & 0.31 &   87 & 0.30 &    21\\
807   &       Brucato & 0.30 &  263 & 0.27 &   220\\
901   &    Euphrosyne & 0.30 & 1476 & 0.28 &  1097\\
902   &        Alauda & 0.28 &  977 & 0.27 &   858\\
904   &       Luthera & 0.32 &  133 & 0.27 &   110\\
905   &       Armenia &  n/a &  n/a & 0.25 &    24\\
\label{tab.w}
\end{longtable}
}


\begin{thebibliography}{XXX}

\bibitem[Bowell \etal(1989)]{bowell89}
Bowell, E., Hapke, B., Domingue, D., Lumme, K., Peltoniemi, J. \& Harris, A.W., 1989, Asteroids II, University of Arizona Press, 524.

\bibitem[Cellino \etal(2002)]{cellino02}
  Cellino, A., Bus, S.J., Doressoundiram, A., Lazzaro, D., Asteroids III, W. F. Bottke Jr., A. Cellino, P. Paolicchi, and R. P. Binzel (eds), University of Arizona Press, 633.

\bibitem[Harris \& Lagerros(2002)]{harris02}
  Harris, A.W. \& Lagerros, J.S.V., Asteroids III, W. F. Bottke Jr., A. Cellino, P. Paolicchi, and R. P. Binzel (eds), University of Arizona Press, 205.

\bibitem[Mainzer \etal(2011a)]{mainzer11}
  Mainzer, A.K., Bauer, J.M., Grav, T., Masiero, J., Cutri, R.M., Dailey, J., Eisenhardt, P., McMillan, R.M. \etal, 2011a, ApJ, 731, 53.
    
\bibitem[Mainzer \etal(2011b)]{mainzer11cal}
  Mainzer, A.K., Grav, T., Masiero, J., Bauer, J.M., Wright, E., Cutri, R.M., McMillan, R.S., Cohen, M., Ressler, M., Eisenhardt, P., 2011b, ApJ, 736, 100.

  \bibitem[Mainzer \etal(2011c)]{mainzer11tax}
Mainzer, A.K., Grav, T., Masiero, J., \etal, 2011c, ApJ, 741, 90.


\bibitem[Mainzer \etal(2016)]{mainzer16}
Mainzer, A.K., Bauer, J., Cutri, R., Grav, T., Kramer, E., \etal, 2016, NASA Planetary Data System, EAR-A-COMPIL-5-NEOWISEDIAM-V1.0.

\bibitem[Masiero \etal(2011)]{masiero11}
  Masiero, J.R., Mainzer, A.K., Grav, T., Bauer, J.M., \etal, 2011, ApJ, 741, 68.

\bibitem[Masiero \etal(2013)]{masiero13}
  Masiero, J.R., Mainzer, A., Bauer, J., Grav, T., Nugent, C., Stevenson, R., 2013, ApJ, 770, 7.
  
\bibitem[Masiero \etal(2012)]{masiero14}
  Masiero, J.R., Grav, T., Mainzer, A.K., Nugent, C., Bauer, J., Stevenson, R., Sonnett, S., 2014, ApJ, 791, 121.

\bibitem[Masiero \etal(2015)]{masieroAIV}
  Masiero, J.R., DeMeo, F.E., Kasuga, T., Parker, A.H., 2015, Asteroids IV, P. Michel, F.E. DeMeo, W.F. Bottke (eds), University of Arizona Press, 323.

\bibitem[Michel \etal(2015)]{michelAIV}
  Michel, P., Richardson, D., Durda, D., Jutzi, M., Asphaug, E., 2015, Asteroids IV, P. Michel, F.E. DeMeo, W.F. Bottke (eds), University of Arizona Press, 341.

\bibitem[Milani \etal(2014)]{milani14}
  Milani, A., Cellino, A., Kne\v{z}evi\'{c}, Z., Novakovi\'{c}, B., Spoto, F., Paolicchi, P., 2014, Icarus, 239, 46.
  
\bibitem[Nesvorn\'{y} \etal(2015)]{nesvornyAIV}
  Nesvorn\'{y}, D., Bro\v{z}, M., Carruba, V., 2015, Asteroids IV, P. Michel, F.E. DeMeo, W.F. Bottke (eds), University of Arizona Press, 297.

\bibitem[Nesvorn\'{y}(2015)]{nesvornyPDS}
  Nesvorn\'{y}, D., 2015, EAR-A-VARGBDET-5-NESVORNYFAM-V3.0, NASA Planetary Data System.
  
\bibitem[Pravec \etal(2012)]{pravec12}
  Pravec, P., Harris, A.W., Ku\v{s}nir\'{a}k, P., Gal\'{a}d, A. \& Hornoch, K., 2012, Icarus, 221, 365.

\bibitem[Vere\v{s} \etal(2015)]{veres15}
  Vere\v{s}, P., Jedicke, R., Fitzsimmons, A., Denneau, L., Granvik, M. \etal, 2015, Icarus, 261, 34.

  
\end{thebibliography}
\end{document}